\documentclass[a4paper]{article}

\begin{document}

\begin{flushright}
\hfill CAMS/99-03\\
%\hfill hep-th/9907037\\
\end{flushright}
\vspace{1cm}

\begin{center}
\baselineskip=16pt {\large \ }{\Large \textbf{Supersymmetric D0-Branes in
Curved Backgrounds }}\vskip2cm \textbf{Ali H. Chamseddine} \\[0pt]
\vskip 0.5cm \emph{Center for Advanced Mathematical Sciences\\[0pt]
and\\[0pt]
Physics Department\\[0pt]
American University of Beirut \\[0pt]
Beirut, Lebanon}\\[0pt]
chams@aub.edu.lb \vskip0.2cm
\end{center}

\vskip 1cm
\begin{abstract}
An action for supersymmetric D0-branes in curved backgrounds is
obtained by dimensional reduction of N=1 ten-dimensional
supergravity coupled to super Yang-Mills system to 0+1 dimensions.
The resultant action exhibits the coset-space symmetry
$\frac{SO(9,9+n)}{SO(9)\times SO(9+n)}\times U(1)$ where
$n=N^{2}-1$ is the dimension of the SU(N) gauge group.
\end{abstract}
\bigskip \bigskip \newpage

\section{Introduction}

Very little is known about D-branes \cite{polchinski}, \cite{A. Sen}, \cite
{susskind}, in curved backgrounds \cite{Douglas} and what restrictions, if
any, should be imposed on such backgrounds . Supersymmetry imposes
constraints on the background metric. In the case of the superstring with
world sheet supersymmetry, there is a direct relation between the number of
supersymmetries and the background metric \cite{Frohlich}. In flat
background geometry the D0-brane action has 16 space-time supersymmetries,
and it is natural to ask for the type of curved backgrounds compatible with
this symmetry. The action with 8 space-time supersymmetries ($N=2$ in 4
dimensions ) was shown to correspond to K\"{a}hler backgrounds \cite{Kato}.

In general it is difficult to construct such actions without
having determined the underlying symmetry. There are few possible
routes to handle this problem, the most obvious one is to quantize
the D-brane action in the presence of a general superspace metric,
thus keeping all supersymmetries, and to determine what are the
required constraints on such a metric \cite {Sezgin},
\cite{hoppe}. This is expected to be extremely complicated and
only recently some work has been done in this direction
\cite{wati}. The other possibility is to determine the necessary
fields to make a supersymmetric multiplet and to find the
corresponding invariant action under such transformations. We
shall follow a simpler approach, which is straightforward but
which the drawback that the underlying symmetry is not manifest.
The idea is based on the observation that the supersymmetric D-0
brane action was obtained by dimensionally reducing the super
Yang-Mills action from ten to 0+1 dimensions \cite{hoppe}. The
most general supersymmetric interaction in ten dimensions with
$N=1$ supersymmetry is that of supergravity coupled to super
Yang-Mills with an arbitrary gauge group \cite{chams}. Dimensional
reduction keeps maximal supresymmetry and rearranges the scalar
fields to have the action of a non-linear sigma model on a coset
space. Reducing to 0+1 dimensions have the peculiarity that there
is no gravitational part and the only fields coming from the
gravity sector in ten-dimensions yields scalar fields. Similarly
the vector fields coming from the Yang-Mills part give scalar
fields taking values in the adjoint representation of the gauge
group. As with compactification of supergravity theory it is
expected that the coset space to be of the form \cite{coset}
\[
\frac{SO(9,9+n)}{SO(9)\times SO(9+n)}\times U(1)
\]
where $n=N^{2}-1$ is the dimension of $SU(N)$ gauge group. T\-he absence of
the gravitational sector in 0+1 dimensions also enables us to identify the
gravitational coupling in higher dimensions with the string tension to
insure that all rescaled scalar fields have the same dimensions and could
serve as coordinates of the D-brane.

The aim of this letter is to derive the D0-brane action with the most
general curved background compatible with maximal space-time supersymmetry.
This is done by compactifying the ten-dimensional theory and grouping all
the resultant fields. The result obtained does not have manifest symmetry.
Nonetheless, this suggests that a direct derivation in terms of
supermultiplets, where some auxiliary fields as well as constrained
variables are used, might drastically simplify the answer. This is the
situation encountered in the derivation of the four-dimensional $N=4$
supersymmetric action where superconformal methods were used to simplify the
analysis \cite{thesis}. This more systematic approach will be left for the
future and our study here will be limited to the action obtained by
dimensional reduction. The plan of this paper is \ as follows. In section 2
we derive the bosonic dimensionally reduced action and in section 3 we give
the fermionic part. Section four includes comments on the results.

\section{Bosonic action of curved D0-branes.}

Our starting point is the $N=1$ supergravity Lagrangian in ten dimensions
coupled to super Yang-Mills system. This is given by (up to quartic
fermionic terms) \cite{chams}
\begin{eqnarray*}
\det \left( e_{M}^{A}\right) ^{-1}L &=&-\frac{1}{4\kappa ^{2}}R(\omega )-%
\frac{i}{2}\overline{\psi }_{M}\Gamma ^{MNP}D_{N}\psi _{P}+\frac{1}{2\kappa
^{2}}\partial _{M}\phi \partial _{N}\phi g^{MN} \\
&&+\frac{i}{2}\overline{\chi }\Gamma ^{M}D_{M}\chi +\frac{1}{\sqrt{2}}%
\partial _{M}\phi \overline{\psi }_{M}\chi +e^{-2\phi }F_{MNP}^{\prime
}F_{QRS}^{\prime }g^{MQ}g^{NR}g^{PS} \\
&&+\frac{i\kappa }{24}e^{-\phi }\overline{\psi }_{M}\left( \Gamma
^{MNPQR}-6g^{MP}\Gamma ^{Q}g^{RN}\right) \psi _{N}F_{PQR}^{\prime } \\
&&-\frac{1}{4}e^{-\phi }Tr\left( G_{MN}G_{PQ}\right) g^{MP}g^{NQ}+\frac{i}{2}%
Tr\left( \overline{\lambda }\Gamma ^{M}D_{M}\lambda \right) \\
&&-\frac{i\kappa }{2\sqrt{2}}Tr\left( \overline{\lambda }\Gamma ^{M}\Gamma
^{NP}\psi _{M}G_{NP}\right)
\end{eqnarray*}
where $F_{MNP\ }^{\prime }$is the field \ strength of the antisymmetric
tensor $B_{MN}$ modified by the gauge Chern-Simons three form.
\[
F_{MNP}^{\prime }=F_{MNP}+\omega _{MNP}^{\left( CS\right) }
\]
and $F_{MNP}=\frac{3}{\kappa }\partial _{\left[ M\right. }B_{NP\left.
{}\right] }$, while
\[
\omega _{MNP}^{\left( CS\right) }=6\kappa Tr\left( A_{\left[ M\right.
}\partial _{N}A_{P\left. {}\right] }+\frac{2}{3}A_{\left[ M\right.
}A_{N}A_{\left. P\right] }\right)
\]
We can reduce this action from $10$ to $d$ dimensions with the following
distribution of fields. The metric $g_{MN}$ gives a metric $g_{\mu \nu }$, $%
m $ vectors and $\frac{1}{2}m\left( m+1\right) $ scalars in $d$ dimensions,
where $m=10-d$. The antisymmetric tensor $B_{MN}$ gives $B_{\mu \nu }$, $m$
vectors and $\frac{1}{2}m\left( m-1\right) $ scalars. The gauge fields $%
A_{M}^{i}$ give $A_{\mu }^{i}$, and $nm$ scalars $A_{m}^{i}$, where $n$ is
the dimension of the gauge group. All in all we will have $m\left(
n+m\right) $ scalars which will span the coset space \cite{coset}
\[
\frac{SO\left( m,n+m\right) }{SO\left( m\right) \times SO\left( n+m\right) }
\]
The case when $d=4$ (i.e. $m=6$) is well established \cite{chams}. The case
we are interested in have $\ m=9$.

To reduce this action to 0+1 dimensions we decompose:
\[
e_{M}^{A}=\left(
\begin{array}{cc}
e_{\stackrel{.}{0}}^{0} & B_{\stackrel{.}{0}}^{a} \\
0 & e_{m}^{a}
\end{array}
\right)
\]
The inverse metric is
\[
e_{A}^{M}=\left(
\begin{array}{cc}
e_{0}^{\stackrel{.}{0}} & e_{0}^{m} \\
0 & e_{a}^{m}
\end{array}
\right)
\]
where $e_{0}^{\stackrel{.}{0}}=\left( e_{\stackrel{.}{0}}^{0}\right) ^{-1}$,
$e_{a}^{m}e_{m}^{b}=\delta _{a}^{b}$, $e_{0}^{m}=-e_{0}^{\stackrel{.}{0}}B_{%
\stackrel{.}{0}}^{a}e_{a}^{m}.$

To evaluate $-\frac{1}{4}\det \left( e_{M}^{A}\right) R(\omega )$ we have
\cite{cremmer}
\[
\frac{1}{4}\det \left( e_{M}^{A}\right) R(\omega )=\frac{1}{16}e\left(
\Omega _{ABC}\Omega ^{ABC}-2\Omega _{ABC}\Omega ^{CAB}-4\Omega _{CA}^{\quad
A}\Omega _{\quad B}^{CB}\right)
\]
where $\Omega _{ABC}=-e_{A}^{M}e_{B}^{N}(\partial _{M}e_{N}^{C}-\partial
_{N}e_{M}^{C})$. The only non-vanishing $\Omega _{ABC}$ is
\[
\Omega _{0bc}=-e_{0}^{\stackrel{.}{0}}e_{b}^{n}\partial _{\stackrel{.}{0}%
}e_{nc}
\]
Substituting this gives
\[
-\frac{1}{4}\det \left( e_{M}^{A}\right) R(\omega )=\frac{e}{8}\left( -\frac{%
1}{2}\partial _{\stackrel{.}{0}}g_{mn}\partial _{\stackrel{.}{0}%
}g^{mn}-2\left( \partial _{\stackrel{.}{0}}\ln e\right) ^{2}\right)
\]
where $e=\det (e_{m}^{a})$. The antisymmetric tensor piece gives
\[
\frac{e}{12}e_{\stackrel{.}{0}}^{0}e^{-2\phi }\left( 3F_{ab0}^{\prime
}F_{ab0}^{\prime }-F_{abc}^{\prime }F_{abc}^{\prime }\right)
\]
where we are raising and lowering tangent space indices with the euclidean
metric $\delta _{ab}$, and
\[
F_{abc}^{\prime }=4\kappa e_{a}^{m}e_{b}^{n}e_{c}^{p}Tr\left( A_{\left[
m\right. }A_{n}A_{\left. p\right] }\right)
\]

\[
F_{ab0}^{\prime }=\frac{1}{\kappa }e_{a}^{m}e_{b}^{n}\left( \partial _{%
\stackrel{.}{0}}B_{mn}-\kappa ^{2}Tr\left( A_{m}D_{\stackrel{.}{0}%
}A_{n}-A_{n}D_{\stackrel{.}{0}}A_{m}\right) \right)
\]
where $D_{\stackrel{.}{0}}A_{m}=\partial _{\stackrel{.}{0}}A_{m}+\left[ A_{_{%
\stackrel{.}{0}}}^{\prime },A_{m}\right] $, and $A_{_{\stackrel{.}{0}%
}}^{\prime }=A_{_{\stackrel{.}{0}}}-B_{_{\stackrel{.}{0}}}^{m}A_{m}$. The
redefinition of $A_{_{\stackrel{.}{0}}}$ will insure that the field $B_{_{%
\stackrel{.}{0}}}^{m}$ will not appear in the action and therefore is
irrelevant.

The Yang-Mills part is
\[
-\frac{1}{4}\det \left( e_{M}^{A}\right) e^{-\phi }Tr\left(
G_{MN}G_{PQ}g^{MN}g^{PQ}\right) =-\frac{e}{4}e_{\stackrel{.}{0}}^{0}e^{-\phi
}Tr\left( G_{ab}G_{ab}-2G_{a0}G_{a0}\right)
\]
where
\[
G_{ab}=e_{a}^{m}e_{b}^{n}\left[ A_{m},A_{n}\right]
\]
\[
G_{a0}=-e_{a}^{m}e_{0}^{\stackrel{.}{0}}D_{\stackrel{.}{0}}A_{m}
\]
Grouping all terms together we obtain the bosonic part of the D0-brane
action:
\begin{eqnarray*}
e^{-1}L_{b} &=&\frac{1}{\kappa ^{2}}\left( e_{0}^{\stackrel{.}{0}}\left( -%
\frac{1}{16}\partial _{\stackrel{.}{0}}g_{mn}\partial _{\stackrel{.}{0}%
}g^{mn}-\frac{1}{4}\left( \partial _{\stackrel{.}{0}}\ln e\right) ^{2}+\frac{%
1}{2}\partial _{\stackrel{.}{0}}\phi \partial _{\stackrel{.}{0}}\phi \right.
\right.  \\
&&\qquad \qquad \left. +\frac{1}{4}e^{-2\phi }g^{mp}g^{nq}D_{\stackrel{.}{0}%
}B_{mn}D_{\stackrel{.}{0}}B_{pq}+\frac{\kappa ^{2}}{2}e^{-\phi
}g^{mn}Tr\left( D_{\stackrel{.}{0}}A_{m}D_{\stackrel{.}{0}}A_{n}\right)
\right)  \\
&&\qquad +e_{\stackrel{.}{0}}^{0}\left( -\frac{4\kappa ^{2}}{3}e^{-2\phi
}g^{mq}g^{nr}g^{ps}Tr\left( A_{\left[ m\right. }A_{n}A_{p\left. {}\right]
}\right) Tr\left( A_{\left[ q\right. }A_{r}A_{s\left. {}\right] }\right)
\right.  \\
&&\qquad \qquad \qquad \left. \left. -\frac{1}{4}e^{-\phi
}g^{mq}g^{nr}Tr\left( \left[ A_{m,}A_{n}\right] \right) Tr\left( \left[
A_{q,}A_{r}\right] \right) \right) \right)
\end{eqnarray*}
where $D_{\stackrel{.}{0}}B_{mn}=\partial _{\stackrel{.}{0}}B_{mn}-\kappa
^{2}Tr\left( A_{m}D_{\stackrel{.}{0}}A_{n}-A_{n}D_{\stackrel{.}{0}%
}A_{m}\right) $. To get the correct dimensions we identify the gravitational
coupling $\kappa $ with the string tension $\alpha ^{\prime }$ and redefine
the gauge fields $A_{m}^{i}=\frac{1}{\alpha ^{\prime }}X_{m}^{i},$thus
identifying them with the D0-brane coordinates. Multiplying the Lagrangian
with an overall factor of $\left( \alpha ^{\prime }\right) ^{2}$ gives the
bosonic part of the D0-brane action. The 82 fields $g_{mn}$, $B_{mn}$ and $%
\phi $ are needed, beside the coordinates $X_{m}^{i}$, to provide
coordinates for a D0-brane action with a curved background. The rescaled
bosonic Lagrangian becomes
\begin{eqnarray*}
e^{-1}L_{b} &=&\left( e_{0}^{\stackrel{.}{0}}\left( -\frac{1}{16}\partial _{%
\stackrel{.}{0}}g_{mn}\partial _{\stackrel{.}{0}}g^{mn}-\frac{1}{4}\left(
\partial _{\stackrel{.}{0}}\ln e\right) ^{2}+\frac{1}{2}\partial _{\stackrel{%
.}{0}}\phi \partial _{\stackrel{.}{0}}\phi \right. \right.  \\
&&\qquad \qquad \left. +\frac{1}{4}e^{-2\phi }g^{mp}g^{nq}D_{\stackrel{.}{0}%
}B_{mn}D_{\stackrel{.}{0}}B_{pq}+\frac{\kappa ^{2}}{2}e^{-\phi
}g^{mn}Tr\left( D_{\stackrel{.}{0}}X_{m}D_{\stackrel{.}{0}}X_{n}\right)
\right)  \\
&&\qquad +e_{\stackrel{.}{0}}^{0}\left( -\frac{4}{3\left( \alpha ^{\prime
}\right) ^{2}}e^{-2\phi }g^{mq}g^{nr}g^{ps}Tr\left( X_{\left[ m\right.
}X_{n}X_{p\left. {}\right] }\right) Tr\left( X_{\left[ q\right.
}X_{r}X_{s\left. {}\right] }\right) \right.  \\
&&\qquad \qquad \qquad \left. \left. -\frac{1}{4\left( \alpha ^{\prime
}\right) ^{2}}e^{-\phi }g^{mq}g^{nr}Tr\left( \left[ X_{m,}X_{n}\right]
\right) Tr\left( \left[ X_{q,}X_{r}\right] \right) \right) \right)
\end{eqnarray*}

The scalar fields can be regrouped into a set $X_{m}^{R}$ where $R=1,\cdots
,9+n$ plus an additional scalar field as a combination of the fields $g_{mn}$%
, $B_{mn}$ , $\phi $ and $X_{m}^{i}.$Can one take the limit where $%
g_{mn}=\delta _{mn}$ $,$ $B_{mn}=0$ and $\phi =0$ ? This gives the flat
background D-0 brane action plus the order 6 terms in $X_{m}^{i}.$This is
usually incompatible with supersymmetry as we shall show later. The proper
limit to flat backgrounds can be obtained by keeping the couplings $\kappa $
and $\alpha ^{\prime }$ independent, then taking the limit $\kappa
\rightarrow 0$.

The transformation law for $e_{\stackrel{.}{0}}^{0}$ with respect to time
transformation is given by
\[
\delta e_{\stackrel{.}{0}}^{0}=\partial _{_{\stackrel{.}{0}}}\left( \xi
^{^{_{\stackrel{.}{0}}}}e_{\stackrel{.}{0}}^{0}\right)
\]
which would allow us to set $e_{\stackrel{.}{0}}^{0}=1$.

In this action the coset space symmetry is not manifest. The coset space
metric is a non-polynomial function of $g_{mn}$, $B_{mn}$, $X_{m}^{i}$ and $%
\phi .$ To obtain manifest symmetry, one method would be to start with the
symmetry $SO(9,9+n)$ using supersymmetric multiplets, and then gauge the $%
SO(9)\times SO(9+n)$ subgroup. This will be the topic of a forthcoming
project where a systematic analysis of all possible background symmetries
would be carried out.

\section{The fermionic action}

The Rarita-Schwinger term
\[
-\frac{i}{2}\det (e_{M}^{A})\overline{\psi }_{A}\Gamma ^{ABC}D_{B}\psi _{C}
\]
where $\psi _{A}=e_{A}^{M}\psi _{M}$, and $D_{M}\psi _{N}=\left( \partial
_{M}+\frac{1}{4}\omega _{M}^{\quad AB}\Gamma _{AB}\right) \psi _{N{\ ,}}$%
gives upon compactification
\begin{eqnarray*}
&&-\frac{i}{8}e\left( \overline{\psi }_{a}\Gamma _{b}\psi _{d}\left(
e_{b}^{n}\partial _{\stackrel{.}{0}}e_{nd}+e_{d}^{n}\partial _{\stackrel{.}{0%
}}e_{nb}\right) -2\overline{\psi }_{0}\Gamma ^{c}\psi _{c}\left(
e_{d}^{n}\partial _{\stackrel{.}{0}}e_{nd}\right) \right) \\
&&+\frac{i}{2}e\left( \overline{\psi }_{a}\Gamma ^{ac}\Gamma _{0}\partial
_{_{\stackrel{.}{0}}}\psi _{c}+\frac{1}{4}\overline{\psi }_{a}\Gamma
^{ac}\Gamma _{de}\Gamma _{0}\psi _{c}\left( e_{d}^{n}\partial _{\stackrel{.}{%
0}}e_{ne}\right) \right) \\
&&+\frac{i}{4}ee_{0}^{\stackrel{.}{0}}\left( \overline{\psi }_{a}\Gamma
^{a}\psi _{0}\left( e_{b}^{n}\partial _{\stackrel{.}{0}}e_{nb}\right) +\frac{%
1}{2}\overline{\psi }_{a}\Gamma _{b}\psi _{0}\left( e_{b}^{n}\partial _{%
\stackrel{.}{0}}e_{na}+e_{a}^{n}\partial _{\stackrel{.}{0}}e_{nb}\right)
\right)
\end{eqnarray*}
where we have used $\psi _{0}=e_{0}^{_{_{\stackrel{.}{0}}}}\left( \psi _{_{_{%
\stackrel{.}{0}}}}-B_{_{_{\stackrel{.}{0}}}}^{a}e_{a}^{m}\psi _{m}\right) $
and $\psi _{a}=e_{a}^{m}\psi _{m}$. Again, the definition of $\psi _{0}$
insures that $B_{_{_{\stackrel{.}{0}}}}^{a}$ does not appear in the action.
The nonvanishing components of the spin-connection are
\begin{eqnarray*}
\omega _{0bc} &=&\frac{1}{2}e_{0}^{\stackrel{.}{0}}\left( e_{b}^{n}\partial
_{\stackrel{.}{0}}e_{nc}-e_{c}^{n}\partial _{\stackrel{.}{0}}e_{nb}\right) \\
\omega _{ab0} &=&-\frac{1}{2}e_{0}^{\stackrel{.}{0}}\left( e_{b}^{n}\partial
_{\stackrel{.}{0}}e_{na}+e_{a}^{n}\partial _{\stackrel{.}{0}}e_{nb}\right)
\end{eqnarray*}
The term $\frac{i}{2}\det \left( e_{M}^{A}\right) \overline{\chi }\Gamma
^{A}D_{A}\chi $ reduces to
\[
-\frac{i}{4}e\left( e_{\stackrel{.}{0}}^{0}\overline{\chi }\Gamma ^{0}\chi
\left( e_{a}^{n}\partial _{\stackrel{.}{0}}e_{na}\right) -2\overline{\chi }%
\Gamma ^{0}\left( \partial _{\stackrel{.}{0}}+\frac{1}{4}e_{b}^{n}\partial _{%
\stackrel{.}{0}}e_{nc}\Gamma _{bc}\right) \chi \right)
\]
Similarly the gaugino kinetic term gives upon reduction
\[
-\frac{i}{4}eTr\left( e_{\stackrel{.}{0}}^{0}\overline{\lambda }\Gamma
^{0}\lambda \left( e_{a}^{n}\partial _{\stackrel{.}{0}}e_{na}\right) -2%
\overline{\lambda }\Gamma ^{0}\left( D_{\stackrel{.}{0}}+\frac{1}{4}%
e_{b}^{n}\partial _{\stackrel{.}{0}}e_{nc}\Gamma _{bc}\right) \lambda
\right)
\]

Next, the fermi-bose interaction $\overline{\psi }\psi F^{\prime }$ gives
\begin{eqnarray*}
&&\frac{i\kappa ^{2}}{24}e^{-\phi }ee_{\stackrel{.}{0}}^{0}\left( 4\left(
\overline{\psi }_{a}\Gamma ^{abcde}\psi _{b}+6\overline{\psi }_{0}\Gamma
^{0bcde}\psi _{b}-6\overline{\psi }_{c}\Gamma _{d}\psi _{e}\right)
TrA_{\left[ c\right. }A_{d}A_{\left. e\right] }\right. \\
&&\qquad \qquad \quad +\left( 3\overline{\psi }_{a}\Gamma ^{abcd0}\psi _{b}-2%
\overline{\psi }_{0}\Gamma _{c}\psi _{d}+2\overline{\psi }_{c}\Gamma
_{0}\psi _{d}-2\overline{\psi }_{c}\Gamma _{d}\psi _{0}\right) \\
&&\ \qquad \qquad \quad \left. \times e_{c}^{m}e_{d}^{n}e_{0}^{\stackrel{.}{0%
}}\left( \frac{1}{\kappa }\partial _{\stackrel{.}{0}}B_{mn}-\kappa Tr\left(
A_{m}D_{\stackrel{.}{0}}A_{n}-A_{n}D_{\stackrel{.}{0}}A_{m}\right) \right)
\right)
\end{eqnarray*}
Finally the $\overline{\psi }\chi \partial \phi $ coupling gives
\[
\frac{1}{2}e\partial _{_{\stackrel{.}{0}}}\phi \overline{\psi }^{0}\chi .
\]

The supersymmetry transformations in ten dimensions are
\begin{eqnarray*}
\delta e_{M}^{A} &=&-i\kappa \overline{\epsilon \,}\Gamma ^{A}\psi _{M} \\
\delta \phi &=&-\frac{\kappa }{\sqrt{2}}\overline{\epsilon \,}\chi \\
\delta B_{MN} &=&\kappa e^{\phi }\left( i\overline{\epsilon }\,\Gamma
_{\left[ M\right. }\psi _{\left. N\right] }+\frac{1}{2}\overline{\epsilon }%
\,\Gamma _{MN}\chi \right) \\
\delta \psi _{M} &=&\frac{1}{\kappa }D_{M}\epsilon +\frac{1}{48}e^{-\phi
}\left( \Gamma _{\qquad M}^{NPQ}+9\delta _{M}^{N}\Gamma ^{PQ}\right)
\epsilon \,F_{NPQ}^{\prime } \\
\delta \chi &=&\frac{i}{2\kappa }\Gamma ^{M}\epsilon \,\partial _{M}\phi \\
\delta A_{M} &=&\frac{i}{\sqrt{2}}e^{\phi }\overline{\epsilon \,}\Gamma
_{M}\lambda \\
\delta \lambda &=&e^{-\frac{1}{2}\phi }\Gamma ^{MN}\epsilon \,G_{MN}
\end{eqnarray*}

The compactified supersymmetry transformations become after rescaling
\begin{eqnarray*}
\delta e_{\stackrel{.}{0}}^{0} &=&-i\kappa \overline{\epsilon }\,\Gamma
^{0}\psi _{\stackrel{.}{0}} \\
\delta e_{\stackrel{.}{0}}^{a} &=&-i\kappa \overline{\epsilon }\,\Gamma
^{a}\psi _{\stackrel{.}{0}} \\
\delta e_{m}^{a} &=&-i\overline{\epsilon }\,\Gamma ^{a}\psi _{m} \\
\delta \phi &=&-\frac{\kappa }{\sqrt{2}}\overline{\epsilon }\chi \\
\delta B_{mn} &=&\kappa e^{\phi }\left( i\overline{\epsilon }\,\Gamma
_{\left[ m\right. }\psi _{\left. n\right] }+\frac{1}{2\sqrt{2}}\overline{%
\epsilon }\,\Gamma _{mn}\chi \right) \\
\delta \psi _{0} &=&\frac{1}{\kappa }e_{0}^{\stackrel{.}{0}}\left( \partial
_{\stackrel{.}{0}}+\frac{1}{4}e_{a}^{m}\partial _{\stackrel{.}{0}%
}e_{mb}\Gamma ^{ab}\right) \epsilon \\
&&+\frac{1}{12}e^{-\phi }\Gamma ^{cd}\epsilon \,e_{c}^{m}e_{d}^{n}\left(
\frac{1}{\kappa }\partial _{\stackrel{.}{0}}B_{mn}-\frac{\kappa }{\left(
\alpha ^{\prime }\right) ^{2}}Tr\left( X_{m}D_{\stackrel{.}{0}}X_{n}-X_{n}D_{%
\stackrel{.}{0}}X_{m}\right) \right) \epsilon \\
&&+\frac{\kappa }{12\left( \alpha ^{\prime }\right) ^{3}}e^{-\phi }\Gamma
^{bcd}\Gamma _{0}\epsilon \,e_{b}^{m}e_{c}^{n}e_{d}^{p}Tr\left( X_{\left[
m\right. }X_{n}X_{\left. p\right] }\right) \\
\delta \psi _{a} &=&\frac{1}{4\kappa }e_{0}^{\stackrel{.}{0}}\left(
e_{a}^{n}\partial _{\stackrel{.}{0}}e_{nb}+e_{b}^{n}\partial _{\stackrel{.}{0%
}}e_{na}\right) \Gamma _{b0}\epsilon \\
&&-\frac{\kappa }{12\left( \alpha ^{\prime }\right) ^{3}}e^{-\phi }\left(
\Gamma _{abcd}-4\delta _{a}^{b}\Gamma _{cd}\right) \epsilon
\,e_{b}^{m}e_{c}^{n}e_{d}^{p}Tr\left( X_{\left[ m\right. }X_{n}X_{\left.
p\right] }\right) \\
&&+\frac{1}{16}e^{-\phi }\left( \Gamma _{0cda}+\frac{4}{3}\Gamma _{0c}\delta
_{d}^{a}\right) \epsilon \,e_{c}^{m}e_{d}^{n}e_{0}^{_{\stackrel{.}{0}%
}}\left( \frac{1}{\kappa }\partial _{\stackrel{.}{0}}B_{mn}-\frac{\kappa }{%
\left( \alpha ^{\prime }\right) ^{3}}Tr\left( X_{m}D_{\stackrel{.}{0}%
}X_{n}-X_{n}D_{\stackrel{.}{0}}X_{m}\right) \right) \\
\delta \chi &=&\frac{i}{\kappa \sqrt{2}}\Gamma ^{0}\epsilon \,e_{0}^{%
\stackrel{.}{0}}\partial _{\stackrel{.}{0}}\phi +\frac{i\kappa }{3\sqrt{2}%
\left( \alpha ^{\prime }\right) ^{3}}e^{-\phi }\Gamma ^{abc}\epsilon
\,e_{a}^{m}e_{b}^{n}e_{c}^{p}Tr\left( X_{\left[ m\right. }X_{n}X_{\left.
p\right] }\right) \\
&&+\frac{i}{4\sqrt{2}}\Gamma ^{ab0}\epsilon \,e_{a}^{m}e_{b}^{n}e_{0}^{_{%
\stackrel{.}{0}}}\left( \frac{1}{\kappa }\partial _{\stackrel{.}{0}}B_{mn}-%
\frac{\kappa }{\left( \alpha ^{\prime }\right) ^{3}}Tr\left( X_{m}D_{%
\stackrel{.}{0}}X_{n}-X_{n}D_{\stackrel{.}{0}}X_{m}\right) \right) \\
\delta A_{\stackrel{.}{0}} &=&\frac{i}{\alpha ^{\prime }\sqrt{2}}e^{\frac{1}{%
2}\phi }\overline{\epsilon }\,\Gamma _{\stackrel{.}{0}}\lambda \\
\delta X_{m} &=&\frac{i\alpha ^{\prime }}{\sqrt{2}}e^{\frac{1}{2}\phi }%
\overline{\epsilon }\,\Gamma _{m}\lambda \\
\delta \lambda &=&\frac{1}{\alpha ^{\prime }\sqrt{2}}e^{-\frac{1}{2}\phi
}\Gamma ^{a0}\epsilon \,e_{a}^{m}e_{0}^{\stackrel{.}{0}}D_{\stackrel{.}{0}%
}X_{m}
\end{eqnarray*}

From these transformations it should be clear that the truncation $%
g_{mn}=\delta _{mn}$, $B_{mn}=0$, $\phi =0$ is not consistent with
supersymmetry because the fields $X_{m}^{i}$, $g_{mn}$, $B_{mn}$ and $\phi $
are now mixed to form $9(9+n)+1$ coordinates for the D-0 brane. A proper way
of going to the flat background limit is to keep $\kappa $ and $\alpha
^{\prime }$ distinct, and then take the limit $\kappa \rightarrow 0$.

\section{Comments}

The D0-brane action with maximal $N=16$ space-time supersymmetry,
derived here have the coset symmetry $\frac{SO(9,9+n)}{SO(9)\times
SO(9+n)}$ which, however, is not manifest. The 81 fields which are
not related to the $SU(N)$ gauge group are essential to provide
curvature for the background. We can say that curved backgrounds
are only possible once $n$ gauge fields are embedded into the
above coset structure. One way to improve on this solution is to
start with the light-cone formulation of the supermembrane in
arbitrary background and find under what conditions the quantized
action simplifies in such a manner as not to involve a square
root. This is a difficult problem and the recent work of weakly
coupling D0-branes to curved backgrounds may help to clarify the
situation \cite{wati}. The action constructed in \cite{wati}
contains higher order time derivatives indicating that higher
derivative terms in  ten-dimensional supergravity should also be
included. It would also be very interesting to find out how the 82
scalar fields arise in this formulation.

Another possibility is to study supersymmetry representations in 0+1
dimensions and form multiplets in complete analogy with the one obtained by
superconformal methods in four dimensions. This would have the advantage of
getting the coset space symmetry in a manifestly invariant way. In addition
this would allow to investigate the general problem of finding the relation
between the required degree of space-time supersymmetry and the nature of
the curved background.


\begin{thebibliography}{99}
\bibitem{polchinski}  J. Polchinski, \textit{Phys.Rev. Lett.} \textbf{75 }%
(1995) 4724.

\bibitem{A. Sen}  A. Sen, \textit{Adv. Theor. Math. Phys.} \textbf{2} (1998)
51.

\bibitem{susskind}  T. Banks, W. Fischler, S. H. Shenker and L. Susskind,
\textit{Phys. Rev}. \textbf{D55} (1997) 5112.

\bibitem{Douglas}  M. R. Douglas, \textit{Nucl. Phys. Proc. Suppl. }\textbf{%
68 }(1998) 381.

\bibitem{Frohlich}  J. Frohlich, O. Grandjean and A. Recknagel, \textit{%
Comm. Math. Phys. }\textbf{193 }(1998) 527.

\bibitem{Kato}  M. R. Douglas, A. Kato, H. Ooguri, \textit{Adv. Theor. Math.
Phys.} \textbf{1} (1998) 237.

\bibitem{Sezgin}  E. Bergshoeff, E. Sezgin and P. K. Townsend, \textit{Phys.
Lett. }\textbf{B189 }(1987) 95.

\bibitem{hoppe}  B. de Wit, J. Hoppe and H. Nicolai, \textit{Nucl. Phys. }%
\textbf{B305 }(1988) 545.

\bibitem{wati}  W. Taylor and M. Van Raamsdonk,
\texttt{hep-th/9904095}.
\bibitem{chams}  A. H. Chamseddine, \textit{Nucl. Phys. }\textbf{B185 }%
(1981) 403; \textit{Phys. Rev. }\textbf{D29 }(1981) 3065;

E. Bergshoeff, M. de Roo, B. de Wit and P. van Nieuwenhuizen, \textit{Nucl.
Phys. }\textbf{B195} (1982) 97:

G. Chapline and N. Manton, \textit{Phys. Lett. }\textbf{120B} (1983) 105.

\bibitem{coset}  J.-P. Derendinger and S. Ferrara, in \textit{Supersymmetry
and Supergravity 84 }editors B. de Wit, P. Fayet and P. van Nieuwenhuizen,
World Scientific p. 159, 1984.

\bibitem{thesis}  M. de Roo and P. Wagemans, \textit{Nucl. Phys. }\textbf{%
B262 }(1995) 644.

\bibitem{cremmer}  E. Cremmer and B. Julia, \textit{Nucl. Phys. }\textbf{%
B159 }(1979) 141.
\end{thebibliography}
\end{document}